\documentclass[a4paper,12pt,preprint,aps,floatfix]{revtex4}
\usepackage[utf8]{inputenc}
\usepackage{longtable}
\usepackage{natbib}
\usepackage{graphicx}     
\usepackage{lscape}
\usepackage{color}
\usepackage{amsmath}
\usepackage{array}
\usepackage{pdflscape}

\begin{document}

\newcommand{\hilight}[1]{\colorbox{yellow}{#1}}
\newcommand{\etal}{{\it et al.} }
\newcommand{\ai}{{\it ab initio }}
\newcommand{\vi}{{\it vide infra}}
\newcommand{\cm}{cm$^{-1}$}
\newcommand{\BODC}{DBOC}
\newcommand{\ownPES}{XBRD}
\newcommand{\Hartree}{{\it E}$_{\rm h}$}
\newcommand{\red}[1]{{\color{red} #1}}

\newcommand{\hp}{H$_3^+$}
\newcommand{\htwo}{H$_2$}

\newcommand{\hwat}{H$_2$$^{16}$O}
\def\dboc{{DBOC}}
\def\a0{{$a_{\rm 0}$}} 
\def\JCP{{\it J. Chem. Phys.}}
\def\JPC{{\it J. Phys. Chem.}}
\def\MP{{\it Mol. Phys.}}
\def\CPL{{\it Chem. Phys. Lett.}}
\def\JMSp{{\it J. Mol. Spectrosc.}}
\renewcommand{\baselinestretch}{1.5}


\title{Sub-percent accuracy for the intensity of a near infrared water line at 10670 cm$^{-1}$: experiment and analysis}

\author{
Tom M. Rubin$^{1}$, Marian Sarrazin$^{1}$,  Nikolai F. Zobov$^{2}$,  Jonathan Tennyson$^{3}$, Oleg L. Polyansky$^{3,2,*}$} 

\affiliation{$^{1}$ PTB (Physikalisch-Technische Bundesanstalt), Abbestraße 5, 10587 Berlin, Germany.}
\affiliation{$^{2}$Institute of Applied Physics, Russian Academy of Sciences,
Ulyanov Street 46, Nizhny Novgorod, Russia 603950.}
\affiliation{$^{3}$Department of Physics and Astronomy, University College London, Gower Street,  London WC1E~6BT, UK.}
\email{o.polyansky@ucl.ac.uk}
\label{firstpage}

\date{\today}

\begin{abstract}
 Laser measurements of the intensity of (201) 3$_{22}$ -- (000) 2$_{21}$ near-infrared water
 absorption line at 10670.1 \cm\ are made using three different Herriott cells. These measurements determine the line intensity with an standard deviation below of 0.3~\% by consideration of the new geometrically derived formula for the optical path length without approximations. This determination together with the current accepted value leads to an overall uncertainty of 0.7~\% of the experimentally assessed line intensity which is compared with previous {\it ab initio} predictions. It is found that steady improvements in the both the dipole moment surface (DMS)
 and potential energy surface (PES) used in the theoretical studies leads to systematic better
 agreement with the observation, with the most recent prediction agreeing closely with the experiment.
\end{abstract}

\maketitle

\section{Introduction}
 All molecules can be arguably divided into three unequal categories.
 To the first group belong two-electron systems, such as  \htwo, 
 HHe$^+$ and \hp. The second group comprises 10-electron systems which
 includes HF, water, ammonia and methane as well as H$_3$O$^+$, NH$_4^+$ and CH$_5^+$. 
 The third group consists of the great multitude of remaining molecules.
 Why do we make such an unequal distribution? This division of species is based on the relative
  simplicity and importance of the species considered.

 The fundamental works of Wolniewicz 
 \cite{Wolniewicz1993.h3p, Wolniewicz1995.h3p} 
 on the \htwo\ molecule
 represent the beginning of solving the \ai\ electronic structure problem for two-electron systems.
 Modern developments built on work  
 by Wolniewicz \cite{Wolniewicz1994.h3p, Alijah1995.h3p, Alijah1995a.h3p}
 includes the papers by Pachucki,  Komasa and co-workers 
 \cite{Pachucki2009.h3p, Komasa2011.h3p, Rychlewski1994.h3p},
 which  demonstrate excellent continuation of Wolniewicz's earlier
 work. Indeed, it has been said  \cite{Mielke2003.h3p, Jungen2009.h3p, Varju2011.h3p, Chen2012.h3p} that
 state-of-the-art \ai\ calculations \cite{jt236,jt282} represent a solution  for the \hp\ molecular
 ion; although there are a number of studies showing that further work is required on this important \cite{jt800} and
 fundamental ion \cite{jt512,jt526,jt566,jt581,15AlFrTy.H3+,18MaRoVi.H3+,18MuMaRe.H3+,19MuMaRa.H3+,20JaLexx.H3+}.

 Ten-electron systems have a particular significance 
 because of the importance of many of them in the atmosphere of the Earth, solar system planets and exoplanets.
 Clearly the \ai\ solution for molecules, belonging to this second
 group require greater computational efforts and remains much further
 from a satisfactory or final solution, than the molecules belonging 
 to the two-electron group. The \ai\ predictions of the ro-vibrational 
 energy levels of water reached the 1 \cm\
 level in \cite{jt309} and
 0.1 \cm\ for stretching states in \cite{jt550}, rising to
 0.3 \cm\ when highly excited bending modes are considered.
 However, there are still many improvements need to reach
 the level of accuracy achieved by Wolniewicz for \htwo\ calculations 
 \cite{Wolniewicz1993.h3p, Wolniewicz1995.h3p}.

 Recent focus has turned towards developing theoretical models which give
 accurate predictions for the transition intensities, see \cite{jt522} for example.
 These predictions are important because experimental determinations of the many lines
 required for atmospheric and other models are often only accurate to a few percent.
 Furthermore, precision intensity measurements are suggesting novel uses for the spectroscopy
 of molecules \cite{17Jousten.CO2}.

 However, tests of accurate  ro-vibrational transition intensities require measurements
 with corresponding or even better accuracy.
 The first sub-percent agreement between theory and experiment
 was achieved for  water line intensities in the calculations of Lodi {\it et al.} (LTP) \cite{jt509}. 
 The LTP dipole moment surface (DMS) was used to compute the water line intensities for
 comparisons with the first sub-percent accurate measurements by
 Hodges and Lisak \cite{09LiHaHo.H2O}. 
 However, subsequent measurements of line intensities for some other water bands \cite{15SiHo.H2O}
 disagreed  with the LTP predictions by up to 5 \%. Further improvements in the calculations resulted
 in the sub-percent agreement with these newly measured line intensities
 \cite{jt744}. 
 
 A detailed comparison between theoretical predictions and high accuracy measurements 
 by Birk {\it et al.} \cite{jt687} suggested that while agreement for some
 bands was satisfactory (about 1\%), this was not true for all bands.
 In particular Birk {\it et al.} found that  for high overtone transitions
 in the near infrared or optical theory only agreed with their near IR Fourier Transform
 Spectrometer (FTS) measurements with an accuracy of about 2 \%. 
 Since then further improvements have been achieved in both the theoretical techniques \cite{jt785} 
 and the corresponding calculations \cite{jt744,jt803}. Such improvements
 need to be tested against ever more accurate experimental results and it
 is such a test that we present here.

 In this paper we present 3 experimental determinations of the intensity of near
 IR overtone line of \hwat. These measurements are compared to the generally adopted value, from the FTS measurements of Birk {\it et al.} \cite{jt687} leading to a sub per cent uncertainty of the experimentally assessed intensity. The experimental results are also compared to various theoretical predictions.

\section{Experimental set-up}\label{setup}
\begin{figure}[htbp]
        \centering
        \includegraphics[width=5.0in]{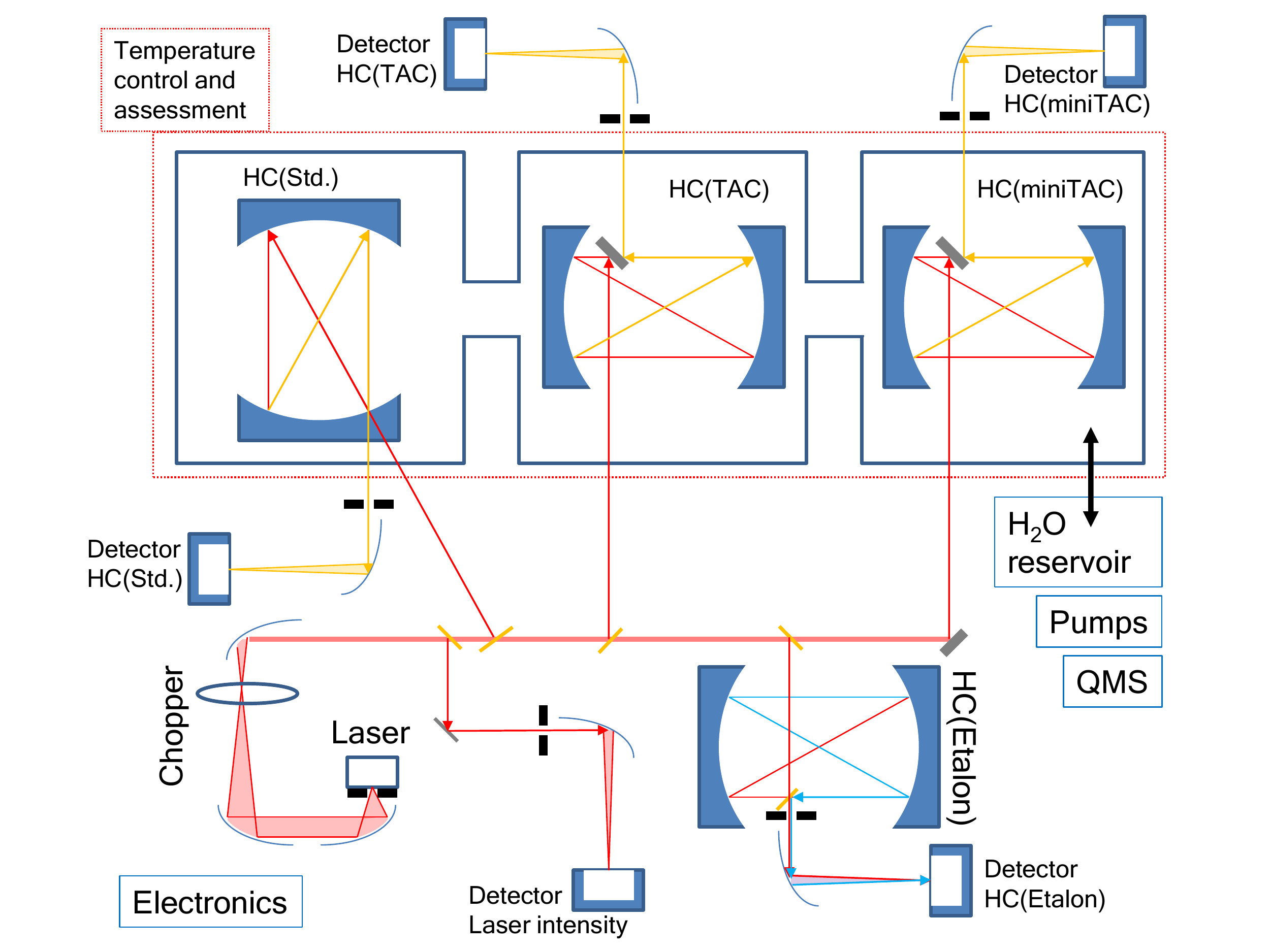}
        \caption{A schematic drawing of the experimental Setup.}
        \label{fig:Setup}
\end{figure}

Figure~\ref{fig:Setup} gives a schematic over of the experimental setup use to make high accuracy measuresments  intensity of the H$_2$$^{16}$O (201) 3$_{22}$ -- (000) 2$_{21}$ line.

A tunable diode laser (Diode: Laser Components; controller: Stanford Research Systems LC501) was used to generate the IR light at 10670 cm$^{-1}$. To suppress interfering influences, such as caused by stray light, the laser beam was chopped at 2 kHz, while the five detector signals were each recorded phase-sensitive with lock-in amplifiers (Stanford Research Systems SR850). All MCT-detectors used were of the same type (Teledyne Judson HgCdTe Photodiode: J23TE2-66C-R01M). The linearity of all detectors was measured extensively beforehand with a set of neutral density filters, and no deviation was detectable even for the maximum laser power used. The possible influence of the detector nonlinearities was therefore considered as negligible for the uncertainty budget, being well below one percent. As illustrated in Fig.~\ref{fig:Setup}), the two partial beams running downward were used to characterize the properties of the laser emission. Here, the first partial beam aimed at 'Detector Laser intensity' is used to determine the intensity fluctuations of the laser and the second partial beam downwards is used for the calibration of the frequency tuning of the laser emission by means of an unbalanced Michelson interferometer in the form of an additional Herriott cell ('HC(Etalon)': FSR = $6.25\times 10^{3}$~cm$^{-1}$).
Each of the three upwards-pointing partial beams  was used to probe its Herriott cell inside the vacuum chamber and was then focused onto the corresponding detector by means of an off-axis paraboloid.

The pressure inside the vacuum chamber was measured with three different gauges, depending on the pressure range: Digiquartz Model 745 from Paroscientific up to 1000~mbar, Baratron from MKS up to 13~mbar and an ionization gauge to assess the residual pressure of $1.4\times10^{-6}$~mbar after 12~h of evacuation. The H$_2$O introduced into the measurement chamber was taken from the headspace of a Mill-Q water supply limiting the maximum H$_2$O pressure to about 13 mbar according to the ambient temperature. The temperature of the vacuum chamber was measured with six PT100 sensors, that were connected to the chamber walls, leading to a total temperature uncertainty of $U(T)<$100~mK (k=1). The gas purity was probed with a mass spectrometer (MKS Microvision-ip1000c).


The experimentally determined H$_2$O line strengths were measured using three different Herriott cells, to increase sensitivity due to the extended optical path length. 
Herriott cells consist of two spherical mirrors with concave surfaces facing each other \cite{herriott1964off, herriott1965folded}. Typically, the coupling of the laser radiation used into and out of cell is done via one or more corresponding holes in the mirrors. In the first cell “$HC_{std}$” both occurred via the same hole in one of the two mirrors. Transverse coupling, for example via a thin plate with reflective surfaces on both sides, is also possible and was used for the cells two: “$HC_{TAC}$” and three: “$HC_{miniTAC}$”.
The optical path length realized inside the Herriott cell for multiple reflections for a complete round trip with $N$ reflections per mirror depends on the distance "$D$" of the mirrors and their effective radii of curvature "$ROC$". The ratio of mirror distance and radius of curvature gives information about the potentially closed configuration and the corresponding number of reflections. The geometrically derived formula:
\begin{eqnarray}
    D = (ROC-s_{corr}) \cdot (1-cos(\pi U/N)) + 2s_{corr} 
    \label{Eq.newHCFormula} 
\end{eqnarray}
describes exactly for which mirror distances closed configurations exist. Here, $U$ describes the number of $2\pi$-twists with respect center line between the two Herriott cell mirrors and $s_{corr}$ is given by $s_{corr}=ROC-\sqrt{ROC^2-r_{sp}^2}$, with $r_{sp}$ as the the radius of the spot patterns generated on the mirrors. More details are given by  Rubin \cite{rubin2017prazise}. The optical path lengths used are specified in Table~I.

\begin{table}[h!]
\caption{Optical path lengths for each Herriott cells (HC); $N$ is the number of reflections and $D$ the
mirror to mirror separation.}
\begin{tabular}{lcccc}
\hline\hline
Name of cell & $D$ & $2N$ & opt. path (inner+outer) & total optical path\\  
\hline
$HC_{std}$ & 0.6664(5)~m & 46 & 30.653(25)~m + 0.260(10)~m & 30.913(27)~m\\
$HC_{TAC}$ & 0.8190(5)~m & 36 & 29.482(45)~m + 0.480(5)~m & 29.962(46)~m\\
$HC_{miniTAC}$ & 0.4038(2)~m & 76 & 30.691(42)~m + 0.266(1)~m & 30.957(42)~m\\
\hline\hline
\end{tabular}
\end{table}

\section{Line intensity measurement results}\label{experimentalresults}
The H$_2$O absorption line strength measurements were made at 296 K. Accordingly, the line strength was determined with the following formula:
\begin{eqnarray}
    S(296 K) = S(T) \cdot (1 - 2.2 \cdot 10^{-3}) \cdot (T - 296 K). 
    \label{Eq.temperature} 
\end{eqnarray}
The H$_2$O pressure range between 0~mbar and 13~mbar was covered.

\begin{table}[h!]
\caption{Measured line intensities at 296~K of the water absorption line at 10670.1 \cm.}
\begin{tabular}{lc}
\hline\hline
Herriott cell & Measured intensity\\  
\hline
$HC_{std}$ & 3.060(36)$\cdot10^{-22}$~cm/molecule\\
$HC_{TAC}$ & 3.074(36)$\cdot10^{-22}$~cm/molecule\\
$HC_{miniTAC}$ & 3.057(36)$\cdot10^{-22}$~cm/molecule\\
\hline\hline
\end{tabular}
\end{table}

Table~III lists the quantities and uncertainty contributions used to estimate the uncertainty in the H$_2$O absorption line strength.
\begin{table}[h!]
\caption{Quantities and uncertainty contributions (with $k=1$) for the measured line strength intensities leading to a relative uncertainty of $u(S)$=1.16~\%.}
\begin{tabular}{lcc}
\hline\hline
Quantity & Value & uncertainty contribution\\
\hline
integrated absorption & na & 1~\%\\
partial pressure & 99,73~\% of 0-13~mBar & 0.5~\% \\
optical path length & 30~m & 0.3~\%\\
Temperature (T) & 296.0~K & $3.4\cdot10^{-4}$\\
second density virial coefficient (B(T)) & -34~cm$^3$/mol & $<10^{-4}$\\
residual pressure & $<1.4\cdot10^{-6}$~mBar & $<10^{-4}$\\
\hline\hline
\end{tabular}
\end{table}

\section{Calculation of line intensities}
There have been a series of efforts to improve predicted intensities
\cite{Langhoff.95.calc,Schwenke.00.calc,Huang.13.calc,Nikitin.13.calc,Li.15.calc,Tyut.19.calc,Chang.19.calc}. 
The studies rely on the accurate solution of the nuclear-motion Schr\"{o}dinger equation 
 the use of high quality potential energy surface
(PESs) and dipole moment surfaces (DMSs) from  \ai\ electronic structure calculations. 
For the DMS,
the calculation of \ai\ values at multiple geometries and subsequent
fitting of these points to the analytical form to reproduce
these points with the typical accuracy of 1x10$^{-5}$ Debye is a standard
procedure. For the PES just the \ai\ calculations of the points and
their fit to the corresponding analytical form is not sufficient.
As shown in \cite{jt714}, the improvement of the PES from
one which can be used to  reproduce the rovibrational energy levels of water with an accuracy of 0.1 \cm\ 
to one giving differences of 0.025 \cm\ can change the values of the calculated intensities
by up to 2 \%. At present, no purely \ai\  PES reproduce the observed
ro-vibrational energy levels of water to better than 0.1 \cm. For this reason  the procedure
of starting from a high quality \ai\ PES and then improving it by fitting to the experimental energy
levels has been widely adopted \cite{01Schwenke.H2O,jt308,jt438,Bubu.11.calc,jt714,jt734,jt803}.
For \hwat, this technique has been used to give  close to experimental accuracy of 0.01 \cm\ \cite{jt714,jt803}. As shown below, such accuracy is important for obtaining
subsequent sub-percent accuracy in line intensity calculations. 

In this work we compare results of three attempts to make high accuracy predictions of \hwat\ line intensities.
The first of these used the "Bubukina" PES of Bubukina {\it et al.} \cite{Bubu.11.calc}  constructed
by fitting to ro-vibrational energy  levels up to 25 000 \cm, which were   reproduced  with a 
standard deviation of  0.022 \cm. The second PES is the improved PES15K 
\cite{jt714} which only fitted to ro-vibrational  energy levels below 15 000 \cm, which were reproduced with an accuracy
of 0.011 \cm. The improved PES15K PES has already been shown to result in a significant
improvement of the calculated intensities  \cite{jt714}. Finally we consider the recently constructed HOT\_WAT PES
of Conway {\it et al.} \cite{jt803} fitted ro-vibrational energies over the entire range of their availability which is
almost up to dissociation \cite{jt795}. We note that  the transition frequency of the
line at 10670.1 \cm\ discussed in the present paper is reproduced best by this
PES, to within  only 0.001 \cm. 

Experience has shown that it is best to use {\it ab initio} DMSs \cite{jt475}. There has been a steady
improvement in both calculation and fitting \ai\ DMSs over the years
 \cite{97PaScxx.H2O,00ScPaxx.H2O,jt424,jt509,jt744}. 
Here we consider the LTP2011 DMS of LTP \cite{jt509} which  is based on a set of 2000 internally-contracted multi-reference configuration  interaction (IC-MRCI) points 
calculated with an aug-cc-pCV6Z basis set as energy derivatives (ED). Relativistic corrections to the dipoles were obtained in
a similar manner by computing the derivatives with respect to the external electric field strength of the mass-velocity,
one-electron Darwin (MVD1) relativistic corrections to the
IC-MRCI energies. This DMS gives sub-percent accuracy for some bands but has been shown to give predictions a few \%\ off
for some bands \cite{jt687}.
Secondly, we consider the CKAPTEN DMS of Conway {\it et al.} \cite{jt744} which  was calculated using a similar procedure  but with relativistic corrections obtained using a
Douglas-Kroll-Hess Hamiltonian to order two (DKH2). The number of points
were significantly increased to about 17~500 and an improved fit function used giving a better overall fit.
For HOT\_WAT and CKAPTEN the average deviation of the predicted intensities for the (201) band  considered here  compared to the measured transition
intensities of Birk {\it et al.} \cite{jt687} is only 0.4 \%.

\section{Comparison with theoretical calculations}

\begin{figure}[htbp]
        \centering
        \includegraphics[width=5.0in]{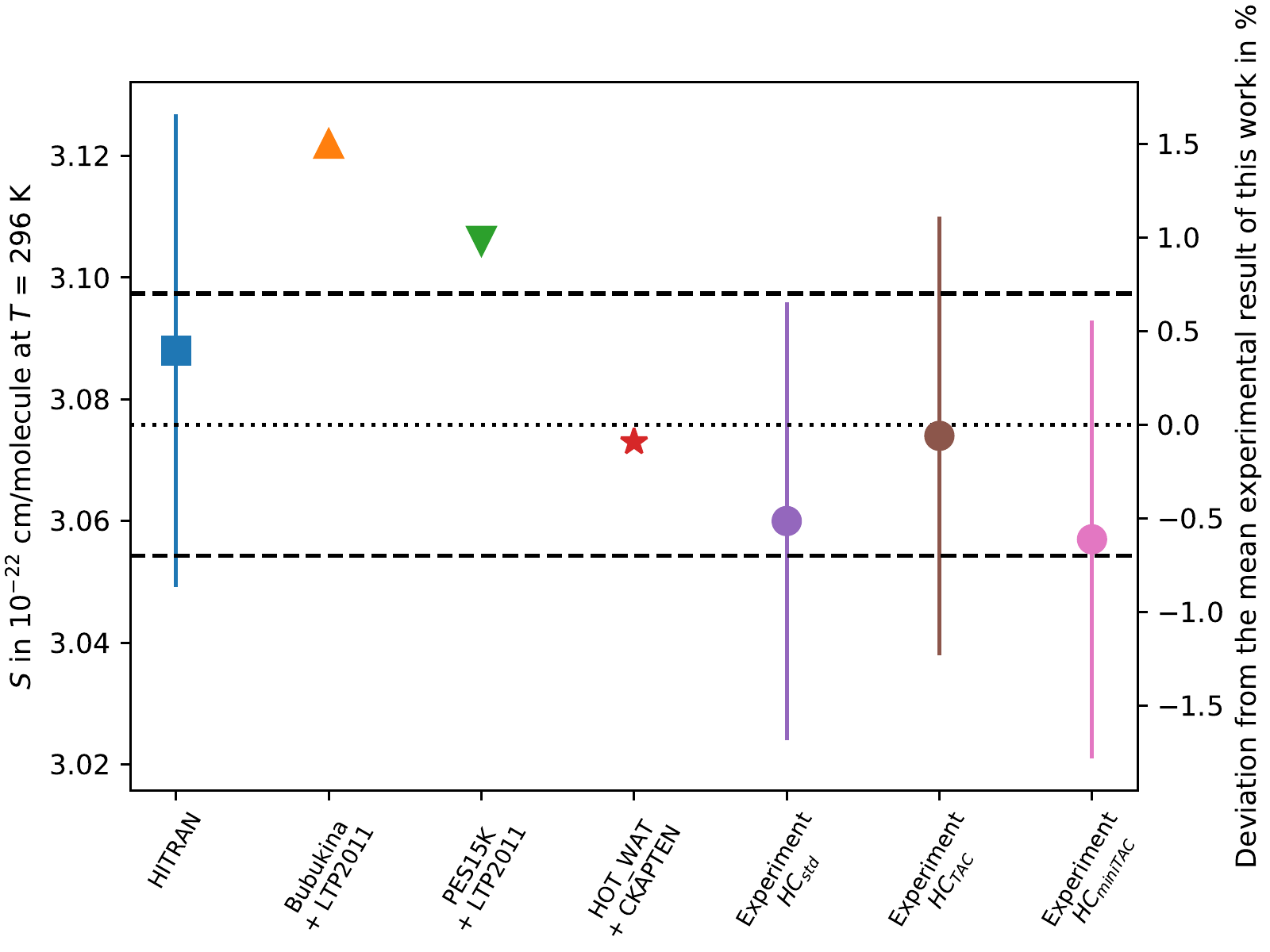}
        \caption{Comparison of measured intensities and theoretical calculations. The dashed horizontal lines represent the mean experimentally assessed line strength of $S$(293 K) = 3.076 $\times 10^{-22}$~cm/molecule with a standard uncertainty of 0.7~\%.}
        \label{fig:compare}
\end{figure}

The comparison between our measurements 
and the theoretical calculations described above
are given in Table \ref{tab:theory} and Fig~\ref{fig:compare}. The first row compares the three
measurements of this work with the value recommended in the HITRAN database \cite{jt836}, 
which actually comes from the FTS measurements of Birk {\it et al.} \cite{jt687} and
has uncertainty of 1~\%. The current measured results considering the new formula for the optical path length have a mean value of 3.064~$\times 10^{-22}$~cm/molecule with a standard deviation of 0.3~\%. Since, most of their uncertainty contributions are considered as "type B" uncertainties with respect to the GUM (Guide to the expression of uncertainty in measurement) the combined value (literature plus the current three results) for the experimentally assessed line strength is 3.076 $\times 10^{-22}$~cm/molecule with a standard uncertainty of 0.7~\%.

The final three rows of Table \ref{tab:theory} compare with theoretical predictions.
The first two of these both use the  LTP2011 DMS \cite{jt714}; agreement improves with
use of the better wavefunctions generated using the more accurate PES15K PES.
The final row compares with the most recent result using both an improved
PES and the CKAPTEN DMS \cite{jt744}. The intensity predicted with these
calculations lies within the experimental uncertainties and differ by less than 0.1\% from the mean measured value. The use of the  CKAPTEN DMS gives a significant improvement over LTP2011.

\begin{table}[h!]
\caption{Comparison of the combined experimental result with the theoretical calculations. Intensities, $S$, at 296~K are give in units of $10^{-22}$~cm/molecule.}
\label{tab:theory}
\begin{tabular}{clcccc}
\hline\hline 
Frequency &$S$&PES &DMS &Exp & o/c   \\
 \cm      & &  &    &     & \%     \\
\hline               
10670.122& 3.088$^a$&  & &3.076& -0.4\\
10670.112& 3.112&Bubukina \cite{Bubu.11.calc}&  LTP2011 \cite{jt509} &3.076& -1.2 \\
10670.115& 3.106&PES15K \cite{jt714}&LTP2011 \cite{jt509}&3.076& -1.0 \\
10670.121& 3.073&HOT\_WAT \cite{jt803}&CKAPTEN \cite{jt744} &3.076& 0.1\\ 
\hline\hline      
\end{tabular}

$^a$ Value from HITRAN \cite{jt836}\\
\end{table}

The first step is very important, though clearly we need the 
expansion in two directions. First, the measurements of intensities
of more lines, belonging to the different vibrational bands.
Secondly, the higher overtones journey towards higher frequencies,
from near IR towards optical region and even UV with the sub-percent
accuracy is necessary.

\section{Conclusions}\label{conclud}
We present new accurate measurements of the intensity of 
the $3_{22} - 2_{21}$ water absorption line in the (201) band at 10670.1 \cm. The combined measurement result gives an intensity $S$(293~K)~=~3.076~$\times~10^{-22}$~cm/molecule with an uncertainty of 0.7\%.
Comparisons with high level theoretical predictions
of this transition intensity show systematically improved agreement as the level
of theoretical treatment of both the dipole moment and potential energy surfaces
are improved. The best calculation gives predictions which lie within the
experimental uncertainty. This comparison suggests that the most recent theory is able to provide
excellent results for higher stretching overtone although further high accuracy experimental
studies would be needed to confirm this situation. The next step is to extend this work
to higher overtones and frequencies extending into the optical region and even the near-UV. 
In this context we note that recent cavity ring down
   spectroscopy measurements 
by Vasilchenko {\it et al.} \cite{21VaMiCa.H2O} suggest that further work is needed to 
get equally reliable predicted intensities for the very weak bending overtones in the red region of spectrum.

\section*{Acknowledgement}
We acknowledge support by State Project IAP RAS No.0030-2021-0016 
and ERC Advanced Investigator Project 883830. T.M.R. and O.L.P. acknowledge support from the QuantumPascal project 18SIB04, which has received funding from the EMPIR programme co-financed by the Participating States and from the European Union’s Horizon 2020 research and innovation programme. 
\newpage


\begin{thebibliography}{58}
\expandafter\ifx\csname natexlab\endcsname\relax\def\natexlab#1{#1}\fi
\expandafter\ifx\csname bibnamefont\endcsname\relax
  \def\bibnamefont#1{#1}\fi
\expandafter\ifx\csname bibfnamefont\endcsname\relax
  \def\bibfnamefont#1{#1}\fi
\expandafter\ifx\csname citenamefont\endcsname\relax
  \def\citenamefont#1{#1}\fi
\expandafter\ifx\csname url\endcsname\relax
  \def\url#1{\texttt{#1}}\fi
\expandafter\ifx\csname urlprefix\endcsname\relax\def\urlprefix{URL }\fi
\providecommand{\bibinfo}[2]{#2}
\providecommand{\eprint}[2][]{\url{#2}}

\bibitem[{\citenamefont{Wolniewicz}(1993)}]{Wolniewicz1993.h3p}
\bibinfo{author}{\bibfnamefont{L.}~\bibnamefont{Wolniewicz}},
  \bibinfo{journal}{JCP} \textbf{\bibinfo{volume}{99}}, \bibinfo{pages}{1851}
  (\bibinfo{year}{1993}).

\bibitem[{\citenamefont{Wolniewicz}(1995)}]{Wolniewicz1995.h3p}
\bibinfo{author}{\bibfnamefont{L.}~\bibnamefont{Wolniewicz}},
  \bibinfo{journal}{JCP} \textbf{\bibinfo{volume}{103}}, \bibinfo{pages}{1792}
  (\bibinfo{year}{1995}).

\bibitem[{\citenamefont{Wolniewicz and Hinze}(1994)}]{Wolniewicz1994.h3p}
\bibinfo{author}{\bibfnamefont{L.}~\bibnamefont{Wolniewicz}} \bibnamefont{and}
  \bibinfo{author}{\bibfnamefont{J.}~\bibnamefont{Hinze}},
  \bibinfo{journal}{JCP} \textbf{\bibinfo{volume}{101}}, \bibinfo{pages}{9817}
  (\bibinfo{year}{1994}).

\bibitem[{\citenamefont{Alijah et~al.}(1995{\natexlab{a}})\citenamefont{Alijah,
  Hinze, and Wolniewicz}}]{Alijah1995.h3p}
\bibinfo{author}{\bibfnamefont{A.}~\bibnamefont{Alijah}},
  \bibinfo{author}{\bibfnamefont{J.}~\bibnamefont{Hinze}}, \bibnamefont{and}
  \bibinfo{author}{\bibfnamefont{L.}~\bibnamefont{Wolniewicz}},
  \bibinfo{journal}{MP} \textbf{\bibinfo{volume}{85}}, \bibinfo{pages}{1105}
  (\bibinfo{year}{1995}{\natexlab{a}}).

\bibitem[{\citenamefont{Alijah et~al.}(1995{\natexlab{b}})\citenamefont{Alijah,
  Wolniewicz, and Hinze}}]{Alijah1995a.h3p}
\bibinfo{author}{\bibfnamefont{A.}~\bibnamefont{Alijah}},
  \bibinfo{author}{\bibfnamefont{L.}~\bibnamefont{Wolniewicz}},
  \bibnamefont{and} \bibinfo{author}{\bibfnamefont{J.}~\bibnamefont{Hinze}},
  \bibinfo{journal}{MP} \textbf{\bibinfo{volume}{85}}, \bibinfo{pages}{1125}
  (\bibinfo{year}{1995}{\natexlab{b}}).

\bibitem[{\citenamefont{Pachucki and Komasa}(2009)}]{Pachucki2009.h3p}
\bibinfo{author}{\bibfnamefont{K.}~\bibnamefont{Pachucki}} \bibnamefont{and}
  \bibinfo{author}{\bibfnamefont{J.}~\bibnamefont{Komasa}},
  \bibinfo{journal}{JCP} \textbf{\bibinfo{volume}{130}},
  \bibinfo{pages}{164113} (\bibinfo{year}{2009}).

\bibitem[{\citenamefont{Komasa et~al.}(2011)\citenamefont{Komasa,
  Piszczatowski, Lach, Przybytek, Jeziorski, and Pachucki}}]{Komasa2011.h3p}
\bibinfo{author}{\bibfnamefont{J.}~\bibnamefont{Komasa}},
  \bibinfo{author}{\bibfnamefont{K.}~\bibnamefont{Piszczatowski}},
  \bibinfo{author}{\bibfnamefont{G.}~\bibnamefont{Lach}},
  \bibinfo{author}{\bibfnamefont{M.}~\bibnamefont{Przybytek}},
  \bibinfo{author}{\bibfnamefont{B.}~\bibnamefont{Jeziorski}},
  \bibnamefont{and} \bibinfo{author}{\bibfnamefont{K.}~\bibnamefont{Pachucki}},
  \bibinfo{journal}{JCTC} \textbf{\bibinfo{volume}{7}}, \bibinfo{pages}{3105}
  (\bibinfo{year}{2011}).

\bibitem[{\citenamefont{Rychlewski et~al.}(1994)\citenamefont{Rychlewski,
  Cencek, and Komasa}}]{Rychlewski1994.h3p}
\bibinfo{author}{\bibfnamefont{J.}~\bibnamefont{Rychlewski}},
  \bibinfo{author}{\bibfnamefont{W.}~\bibnamefont{Cencek}}, \bibnamefont{and}
  \bibinfo{author}{\bibfnamefont{J.}~\bibnamefont{Komasa}},
  \bibinfo{journal}{CPL} \textbf{\bibinfo{volume}{229}},
  \bibinfo{pages}{657–660} (\bibinfo{year}{1994}).

\bibitem[{\citenamefont{Mielke et~al.}(2003)\citenamefont{Mielke, Peterson,
  Schwenke, Garrett, Truhlar, Michael, Su, and Sutherland}}]{Mielke2003.h3p}
\bibinfo{author}{\bibfnamefont{S.}~\bibnamefont{Mielke}},
  \bibinfo{author}{\bibfnamefont{K.}~\bibnamefont{Peterson}},
  \bibinfo{author}{\bibfnamefont{D.}~\bibnamefont{Schwenke}},
  \bibinfo{author}{\bibfnamefont{B.}~\bibnamefont{Garrett}},
  \bibinfo{author}{\bibfnamefont{D.}~\bibnamefont{Truhlar}},
  \bibinfo{author}{\bibfnamefont{J.}~\bibnamefont{Michael}},
  \bibinfo{author}{\bibfnamefont{M.}~\bibnamefont{Su}}, \bibnamefont{and}
  \bibinfo{author}{\bibfnamefont{J.}~\bibnamefont{Sutherland}},
  \bibinfo{journal}{PRL} \textbf{\bibinfo{volume}{91}}, \bibinfo{pages}{063201}
  (\bibinfo{year}{2003}).

\bibitem[{\citenamefont{Jungen and Pratt}(2009)}]{Jungen2009.h3p}
\bibinfo{author}{\bibfnamefont{C.}~\bibnamefont{Jungen}} \bibnamefont{and}
  \bibinfo{author}{\bibfnamefont{S.~T.} \bibnamefont{Pratt}},
  \bibinfo{journal}{PRL} \textbf{\bibinfo{volume}{102}},
  \bibinfo{pages}{023201} (\bibinfo{year}{2009}).

\bibitem[{\citenamefont{Varju et~al.}(2011)\citenamefont{Varju, Hejduk, Dohnal,
  Jilek, Kotrik, Plasil, Gerlich, and Glosik}}]{Varju2011.h3p}
\bibinfo{author}{\bibfnamefont{J.}~\bibnamefont{Varju}},
  \bibinfo{author}{\bibfnamefont{M.}~\bibnamefont{Hejduk}},
  \bibinfo{author}{\bibfnamefont{P.}~\bibnamefont{Dohnal}},
  \bibinfo{author}{\bibfnamefont{M.}~\bibnamefont{Jilek}},
  \bibinfo{author}{\bibfnamefont{T.}~\bibnamefont{Kotrik}},
  \bibinfo{author}{\bibfnamefont{R.}~\bibnamefont{Plasil}},
  \bibinfo{author}{\bibfnamefont{D.}~\bibnamefont{Gerlich}}, \bibnamefont{and}
  \bibinfo{author}{\bibfnamefont{J.}~\bibnamefont{Glosik}},
  \bibinfo{journal}{PRL} \textbf{\bibinfo{volume}{106}},
  \bibinfo{pages}{203201} (\bibinfo{year}{2011}).

\bibitem[{\citenamefont{Chen et~al.}(2012)\citenamefont{Chen, Hsiao, Peng,
  Amano, and Shy}}]{Chen2012.h3p}
\bibinfo{author}{\bibfnamefont{H.}~\bibnamefont{Chen}},
  \bibinfo{author}{\bibfnamefont{C.}~\bibnamefont{Hsiao}},
  \bibinfo{author}{\bibfnamefont{J.}~\bibnamefont{Peng}},
  \bibinfo{author}{\bibfnamefont{T.}~\bibnamefont{Amano}}, \bibnamefont{and}
  \bibinfo{author}{\bibfnamefont{J.}~\bibnamefont{Shy}}, \bibinfo{journal}{PRL}
  \textbf{\bibinfo{volume}{109}}, \bibinfo{pages}{263002}
  (\bibinfo{year}{2012}).

\bibitem[{\citenamefont{Polyansky and Tennyson}(1999)}]{jt236}
\bibinfo{author}{\bibfnamefont{O.~L.} \bibnamefont{Polyansky}}
  \bibnamefont{and} \bibinfo{author}{\bibfnamefont{J.}~\bibnamefont{Tennyson}},
  \bibinfo{journal}{J. Chem. Phys.} \textbf{\bibinfo{volume}{110}},
  \bibinfo{pages}{5056} (\bibinfo{year}{1999}).

\bibitem[{\citenamefont{Tennyson et~al.}(2002)\citenamefont{Tennyson, Barletta,
  Kostin, Polyansky, and Zobov}}]{jt282}
\bibinfo{author}{\bibfnamefont{J.}~\bibnamefont{Tennyson}},
  \bibinfo{author}{\bibfnamefont{P.}~\bibnamefont{Barletta}},
  \bibinfo{author}{\bibfnamefont{M.~A.} \bibnamefont{Kostin}},
  \bibinfo{author}{\bibfnamefont{O.~L.} \bibnamefont{Polyansky}},
  \bibnamefont{and} \bibinfo{author}{\bibfnamefont{N.~F.} \bibnamefont{Zobov}},
  \bibinfo{journal}{Spectrochimica Acta A} \textbf{\bibinfo{volume}{58}},
  \bibinfo{pages}{663} (\bibinfo{year}{2002}).

\bibitem[{\citenamefont{Miller et~al.}(2020)\citenamefont{Miller, Geballe,
  Stallard, and Tennyson}}]{jt800}
\bibinfo{author}{\bibfnamefont{S.}~\bibnamefont{Miller}},
  \bibinfo{author}{\bibfnamefont{T.~R.} \bibnamefont{Geballe}},
  \bibinfo{author}{\bibfnamefont{T.}~\bibnamefont{Stallard}}, \bibnamefont{and}
  \bibinfo{author}{\bibfnamefont{J.}~\bibnamefont{Tennyson}},
  \bibinfo{journal}{Rev. Mod. Phys.} \textbf{\bibinfo{volume}{92}},
  \bibinfo{pages}{035003} (\bibinfo{year}{2020}).

\bibitem[{\citenamefont{Pavanello
  et~al.}(2012{\natexlab{a}})\citenamefont{Pavanello, Adamowicz, Alijah, Zobov,
  Mizus, Polyansky, Tennyson, Szidarovszky, Cs\'asz\'ar, Berg et~al.}}]{jt512}
\bibinfo{author}{\bibfnamefont{M.}~\bibnamefont{Pavanello}},
  \bibinfo{author}{\bibfnamefont{L.}~\bibnamefont{Adamowicz}},
  \bibinfo{author}{\bibfnamefont{A.}~\bibnamefont{Alijah}},
  \bibinfo{author}{\bibfnamefont{N.~F.} \bibnamefont{Zobov}},
  \bibinfo{author}{\bibfnamefont{I.~I.} \bibnamefont{Mizus}},
  \bibinfo{author}{\bibfnamefont{O.~L.} \bibnamefont{Polyansky}},
  \bibinfo{author}{\bibfnamefont{J.}~\bibnamefont{Tennyson}},
  \bibinfo{author}{\bibfnamefont{T.}~\bibnamefont{Szidarovszky}},
  \bibinfo{author}{\bibfnamefont{A.~G.} \bibnamefont{Cs\'asz\'ar}},
  \bibinfo{author}{\bibfnamefont{M.}~\bibnamefont{Berg}}, \bibnamefont{et~al.},
  \bibinfo{journal}{Phys. Rev. Lett.} \textbf{\bibinfo{volume}{108}},
  \bibinfo{pages}{023002} (\bibinfo{year}{2012}{\natexlab{a}}).

\bibitem[{\citenamefont{Pavanello
  et~al.}(2012{\natexlab{b}})\citenamefont{Pavanello, Adamowicz, Alijah, Zobov,
  Mizus, Polyansky, Tennyson, Szidarovszky, and {Cs\'asz\'ar}}}]{jt526}
\bibinfo{author}{\bibfnamefont{M.}~\bibnamefont{Pavanello}},
  \bibinfo{author}{\bibfnamefont{L.}~\bibnamefont{Adamowicz}},
  \bibinfo{author}{\bibfnamefont{A.}~\bibnamefont{Alijah}},
  \bibinfo{author}{\bibfnamefont{N.~F.} \bibnamefont{Zobov}},
  \bibinfo{author}{\bibfnamefont{I.~I.} \bibnamefont{Mizus}},
  \bibinfo{author}{\bibfnamefont{O.~L.} \bibnamefont{Polyansky}},
  \bibinfo{author}{\bibfnamefont{J.}~\bibnamefont{Tennyson}},
  \bibinfo{author}{\bibfnamefont{T.}~\bibnamefont{Szidarovszky}},
  \bibnamefont{and} \bibinfo{author}{\bibfnamefont{A.~G.}
  \bibnamefont{{Cs\'asz\'ar}}}, \bibinfo{journal}{J. Chem. Phys.}
  \textbf{\bibinfo{volume}{136}}, \bibinfo{pages}{184303}
  (\bibinfo{year}{2012}{\natexlab{b}}).

\bibitem[{\citenamefont{Diniz et~al.}(2013)\citenamefont{Diniz, Mohallem,
  Alijah, Pavanello, Adamowicz, Polyansky, and Tennyson}}]{jt566}
\bibinfo{author}{\bibfnamefont{L.~G.} \bibnamefont{Diniz}},
  \bibinfo{author}{\bibfnamefont{J.~R.} \bibnamefont{Mohallem}},
  \bibinfo{author}{\bibfnamefont{A.}~\bibnamefont{Alijah}},
  \bibinfo{author}{\bibfnamefont{M.}~\bibnamefont{Pavanello}},
  \bibinfo{author}{\bibfnamefont{L.}~\bibnamefont{Adamowicz}},
  \bibinfo{author}{\bibfnamefont{O.~L.} \bibnamefont{Polyansky}},
  \bibnamefont{and} \bibinfo{author}{\bibfnamefont{J.}~\bibnamefont{Tennyson}},
  \bibinfo{journal}{Phys. Rev. A} \textbf{\bibinfo{volume}{88}},
  \bibinfo{pages}{032506} (\bibinfo{year}{2013}).

\bibitem[{\citenamefont{Lodi et~al.}(2014)\citenamefont{Lodi, Polyansky,
  J.~Tennyson, and Zobov}}]{jt581}
\bibinfo{author}{\bibfnamefont{L.}~\bibnamefont{Lodi}},
  \bibinfo{author}{\bibfnamefont{O.~L.} \bibnamefont{Polyansky}},
  \bibinfo{author}{\bibfnamefont{A.~A.} \bibnamefont{J.~Tennyson}},
  \bibnamefont{and} \bibinfo{author}{\bibfnamefont{N.~F.} \bibnamefont{Zobov}},
  \bibinfo{journal}{Phys. Rev. A} \textbf{\bibinfo{volume}{89}},
  \bibinfo{pages}{032505} (\bibinfo{year}{2014}).

\bibitem[{\citenamefont{Alijah et~al.}(2015)\citenamefont{Alijah, Fremont, and
  Tyuterev}}]{15AlFrTy.H3+}
\bibinfo{author}{\bibfnamefont{A.}~\bibnamefont{Alijah}},
  \bibinfo{author}{\bibfnamefont{J.}~\bibnamefont{Fremont}}, \bibnamefont{and}
  \bibinfo{author}{\bibfnamefont{V.~G.} \bibnamefont{Tyuterev}},
  \bibinfo{journal}{Phys. Rev. A} \textbf{\bibinfo{volume}{92}},
  \bibinfo{pages}{012704} (\bibinfo{year}{2015}).

\bibitem[{\citenamefont{Marquez-Mijares
  et~al.}(2018)\citenamefont{Marquez-Mijares, Roncero, Villarreal, and
  Gonzalez-Lezana}}]{18MaRoVi.H3+}
\bibinfo{author}{\bibfnamefont{M.}~\bibnamefont{Marquez-Mijares}},
  \bibinfo{author}{\bibfnamefont{O.}~\bibnamefont{Roncero}},
  \bibinfo{author}{\bibfnamefont{P.}~\bibnamefont{Villarreal}},
  \bibnamefont{and}
  \bibinfo{author}{\bibfnamefont{T.}~\bibnamefont{Gonzalez-Lezana}},
  \bibinfo{journal}{Few-Body Syst.} \textbf{\bibinfo{volume}{59}},
  \bibinfo{pages}{14} (\bibinfo{year}{2018}).

\bibitem[{\citenamefont{Muolo et~al.}(2018)\citenamefont{Muolo, Matyus, and
  Reiher}}]{18MuMaRe.H3+}
\bibinfo{author}{\bibfnamefont{A.}~\bibnamefont{Muolo}},
  \bibinfo{author}{\bibfnamefont{E.}~\bibnamefont{Matyus}}, \bibnamefont{and}
  \bibinfo{author}{\bibfnamefont{M.}~\bibnamefont{Reiher}},
  \bibinfo{journal}{J. Chem. Phys.} \textbf{\bibinfo{volume}{149}},
  \bibinfo{pages}{184105} (\bibinfo{year}{2018}).

\bibitem[{\citenamefont{Muolo et~al.}(2019)\citenamefont{Muolo, M{\'a}tyus, and
  Reiher}}]{19MuMaRa.H3+}
\bibinfo{author}{\bibfnamefont{A.}~\bibnamefont{Muolo}},
  \bibinfo{author}{\bibfnamefont{E.}~\bibnamefont{M{\'a}tyus}},
  \bibnamefont{and} \bibinfo{author}{\bibfnamefont{M.}~\bibnamefont{Reiher}},
  \bibinfo{journal}{J. Chem. Phys.} \textbf{\bibinfo{volume}{151}},
  \bibinfo{pages}{154110} (\bibinfo{year}{2019}).

\bibitem[{\citenamefont{Jaquet and Lesiuk}(2020)}]{20JaLexx.H3+}
\bibinfo{author}{\bibfnamefont{R.}~\bibnamefont{Jaquet}} \bibnamefont{and}
  \bibinfo{author}{\bibfnamefont{M.}~\bibnamefont{Lesiuk}},
  \bibinfo{journal}{J. Chem. Phys.} \textbf{\bibinfo{volume}{152}},
  \bibinfo{pages}{104109} (\bibinfo{year}{2020}).

\bibitem[{\citenamefont{Polyansky et~al.}(2003)\citenamefont{Polyansky,
  {Cs\'asz\'ar}, Shirin, Zobov, Barletta, Tennyson, Schwenke, and
  Knowles}}]{jt309}
\bibinfo{author}{\bibfnamefont{O.~L.} \bibnamefont{Polyansky}},
  \bibinfo{author}{\bibfnamefont{A.~G.} \bibnamefont{{Cs\'asz\'ar}}},
  \bibinfo{author}{\bibfnamefont{S.~V.} \bibnamefont{Shirin}},
  \bibinfo{author}{\bibfnamefont{N.~F.} \bibnamefont{Zobov}},
  \bibinfo{author}{\bibfnamefont{P.}~\bibnamefont{Barletta}},
  \bibinfo{author}{\bibfnamefont{J.}~\bibnamefont{Tennyson}},
  \bibinfo{author}{\bibfnamefont{D.~W.} \bibnamefont{Schwenke}},
  \bibnamefont{and} \bibinfo{author}{\bibfnamefont{P.~J.}
  \bibnamefont{Knowles}}, \bibinfo{journal}{Science}
  \textbf{\bibinfo{volume}{299}}, \bibinfo{pages}{539} (\bibinfo{year}{2003}).

\bibitem[{\citenamefont{Polyansky et~al.}(2013)\citenamefont{Polyansky,
  Ovsyannikov, Kyuberis, Lodi, Tennyson, and Zobov}}]{jt550}
\bibinfo{author}{\bibfnamefont{O.~L.} \bibnamefont{Polyansky}},
  \bibinfo{author}{\bibfnamefont{R.~I.} \bibnamefont{Ovsyannikov}},
  \bibinfo{author}{\bibfnamefont{A.~A.} \bibnamefont{Kyuberis}},
  \bibinfo{author}{\bibfnamefont{L.}~\bibnamefont{Lodi}},
  \bibinfo{author}{\bibfnamefont{J.}~\bibnamefont{Tennyson}}, \bibnamefont{and}
  \bibinfo{author}{\bibfnamefont{N.~F.} \bibnamefont{Zobov}},
  \bibinfo{journal}{J. Phys. Chem. A} \textbf{\bibinfo{volume}{117}},
  \bibinfo{pages}{9633–9643} (\bibinfo{year}{2013}).

\bibitem[{\citenamefont{Lodi and Tennyson}(2012)}]{jt522}
\bibinfo{author}{\bibfnamefont{L.}~\bibnamefont{Lodi}} \bibnamefont{and}
  \bibinfo{author}{\bibfnamefont{J.}~\bibnamefont{Tennyson}},
  \bibinfo{journal}{J. Quant. Spectrosc. Radiat. Transf.}
  \textbf{\bibinfo{volume}{113}}, \bibinfo{pages}{850} (\bibinfo{year}{2012}).

\bibitem[{\citenamefont{Jousten et~al.}({2017})\citenamefont{Jousten,
  Hendricks, Barker, Douglas, Eckel, Egan, Fedchak, Flügge, Gaiser, Olson
  et~al.}}]{17Jousten.CO2}
\bibinfo{author}{\bibfnamefont{K.}~\bibnamefont{Jousten}},
  \bibinfo{author}{\bibfnamefont{J.}~\bibnamefont{Hendricks}},
  \bibinfo{author}{\bibfnamefont{D.}~\bibnamefont{Barker}},
  \bibinfo{author}{\bibfnamefont{K.}~\bibnamefont{Douglas}},
  \bibinfo{author}{\bibfnamefont{S.}~\bibnamefont{Eckel}},
  \bibinfo{author}{\bibfnamefont{P.}~\bibnamefont{Egan}},
  \bibinfo{author}{\bibfnamefont{J.}~\bibnamefont{Fedchak}},
  \bibinfo{author}{\bibfnamefont{J.}~\bibnamefont{Flügge}},
  \bibinfo{author}{\bibfnamefont{C.}~\bibnamefont{Gaiser}},
  \bibinfo{author}{\bibfnamefont{D.}~\bibnamefont{Olson}},
  \bibnamefont{et~al.}, \bibinfo{journal}{{Metrologia}}
  \textbf{\bibinfo{volume}{{54}}}, \bibinfo{pages}{S146}
  (\bibinfo{year}{{2017}}).

\bibitem[{\citenamefont{Lodi et~al.}(2011)\citenamefont{Lodi, Tennyson, and
  Polyansky}}]{jt509}
\bibinfo{author}{\bibfnamefont{L.}~\bibnamefont{Lodi}},
  \bibinfo{author}{\bibfnamefont{J.}~\bibnamefont{Tennyson}}, \bibnamefont{and}
  \bibinfo{author}{\bibfnamefont{O.~L.} \bibnamefont{Polyansky}},
  \bibinfo{journal}{J. Chem. Phys.} \textbf{\bibinfo{volume}{135}},
  \bibinfo{pages}{034113} (\bibinfo{year}{2011}).

\bibitem[{\citenamefont{Lisak et~al.}(2009)\citenamefont{Lisak, Havey, and
  Hodges}}]{09LiHaHo.H2O}
\bibinfo{author}{\bibfnamefont{D.}~\bibnamefont{Lisak}},
  \bibinfo{author}{\bibfnamefont{D.~K.} \bibnamefont{Havey}}, \bibnamefont{and}
  \bibinfo{author}{\bibfnamefont{J.~T.} \bibnamefont{Hodges}},
  \bibinfo{journal}{Phys. Rev. A} \textbf{\bibinfo{volume}{79}},
  \bibinfo{pages}{052507} (\bibinfo{year}{2009}).

\bibitem[{\citenamefont{Sironneau and Hodges}(2015)}]{15SiHo.H2O}
\bibinfo{author}{\bibfnamefont{V.~T.} \bibnamefont{Sironneau}}
  \bibnamefont{and} \bibinfo{author}{\bibfnamefont{J.~T.}
  \bibnamefont{Hodges}}, \bibinfo{journal}{J. Quant. Spectrosc. Radiat.
  Transf.} \textbf{\bibinfo{volume}{152}}, \bibinfo{pages}{1}
  (\bibinfo{year}{2015}).

\bibitem[{\citenamefont{Conway et~al.}(2018)\citenamefont{Conway, Kyuberis,
  Polyansky, Tennyson, and Zobov}}]{jt744}
\bibinfo{author}{\bibfnamefont{E.~K.} \bibnamefont{Conway}},
  \bibinfo{author}{\bibfnamefont{A.~A.} \bibnamefont{Kyuberis}},
  \bibinfo{author}{\bibfnamefont{O.~L.} \bibnamefont{Polyansky}},
  \bibinfo{author}{\bibfnamefont{J.}~\bibnamefont{Tennyson}}, \bibnamefont{and}
  \bibinfo{author}{\bibfnamefont{N.}~\bibnamefont{Zobov}}, \bibinfo{journal}{J.
  Chem. Phys.} \textbf{\bibinfo{volume}{149}}, \bibinfo{pages}{084307}
  (\bibinfo{year}{2018}).

\bibitem[{\citenamefont{Birk et~al.}(2017)\citenamefont{Birk, Wagner, Loos,
  Lodi, Polyansky, Kyuberis, Zobov, and Tennyson}}]{jt687}
\bibinfo{author}{\bibfnamefont{M.}~\bibnamefont{Birk}},
  \bibinfo{author}{\bibfnamefont{G.}~\bibnamefont{Wagner}},
  \bibinfo{author}{\bibfnamefont{J.}~\bibnamefont{Loos}},
  \bibinfo{author}{\bibfnamefont{L.}~\bibnamefont{Lodi}},
  \bibinfo{author}{\bibfnamefont{O.~L.} \bibnamefont{Polyansky}},
  \bibinfo{author}{\bibfnamefont{A.~A.} \bibnamefont{Kyuberis}},
  \bibinfo{author}{\bibfnamefont{N.~F.} \bibnamefont{Zobov}}, \bibnamefont{and}
  \bibinfo{author}{\bibfnamefont{J.}~\bibnamefont{Tennyson}},
  \bibinfo{journal}{J. Quant. Spectrosc. Radiat. Transf.}
  \textbf{\bibinfo{volume}{203}}, \bibinfo{pages}{88} (\bibinfo{year}{2017}).

\bibitem[{\citenamefont{Conway et~al.}(2020{\natexlab{a}})\citenamefont{Conway,
  Gordon, Polyansky, and Tennyson}}]{jt785}
\bibinfo{author}{\bibfnamefont{E.~K.} \bibnamefont{Conway}},
  \bibinfo{author}{\bibfnamefont{I.~E.} \bibnamefont{Gordon}},
  \bibinfo{author}{\bibfnamefont{O.~L.} \bibnamefont{Polyansky}},
  \bibnamefont{and} \bibinfo{author}{\bibfnamefont{J.}~\bibnamefont{Tennyson}},
  \bibinfo{journal}{J. Chem. Phys.} \textbf{\bibinfo{volume}{152}},
  \bibinfo{pages}{024105} (\bibinfo{year}{2020}{\natexlab{a}}).

\bibitem[{\citenamefont{Conway et~al.}(2020{\natexlab{b}})\citenamefont{Conway,
  Gordon, Tennyson, Polyansky, Yurchenko, and Chance}}]{jt803}
\bibinfo{author}{\bibfnamefont{E.~K.} \bibnamefont{Conway}},
  \bibinfo{author}{\bibfnamefont{I.~E.} \bibnamefont{Gordon}},
  \bibinfo{author}{\bibfnamefont{J.}~\bibnamefont{Tennyson}},
  \bibinfo{author}{\bibfnamefont{O.~L.} \bibnamefont{Polyansky}},
  \bibinfo{author}{\bibfnamefont{S.~N.} \bibnamefont{Yurchenko}},
  \bibnamefont{and} \bibinfo{author}{\bibfnamefont{K.}~\bibnamefont{Chance}},
  \bibinfo{journal}{Atmos. Chem. Phys.} \textbf{\bibinfo{volume}{20}},
  \bibinfo{pages}{10015} (\bibinfo{year}{2020}{\natexlab{b}}).

\bibitem[{\citenamefont{Herriott et~al.}(1964)\citenamefont{Herriott, Kogelnik,
  and Kompfner}}]{herriott1964off}
\bibinfo{author}{\bibfnamefont{D.}~\bibnamefont{Herriott}},
  \bibinfo{author}{\bibfnamefont{H.}~\bibnamefont{Kogelnik}}, \bibnamefont{and}
  \bibinfo{author}{\bibfnamefont{R.}~\bibnamefont{Kompfner}},
  \bibinfo{journal}{Applied Optics} \textbf{\bibinfo{volume}{3}},
  \bibinfo{pages}{523} (\bibinfo{year}{1964}).

\bibitem[{\citenamefont{Herriott and Schulte}(1965)}]{herriott1965folded}
\bibinfo{author}{\bibfnamefont{D.~R.} \bibnamefont{Herriott}} \bibnamefont{and}
  \bibinfo{author}{\bibfnamefont{H.~J.} \bibnamefont{Schulte}},
  \bibinfo{journal}{Applied Optics} \textbf{\bibinfo{volume}{4}},
  \bibinfo{pages}{883} (\bibinfo{year}{1965}).

\bibitem[{\citenamefont{Rubin}(2017)}]{rubin2017prazise}
\bibinfo{author}{\bibfnamefont{T.~M.} \bibnamefont{Rubin}}, Ph.D. thesis,
  \bibinfo{school}{Freie Universit\"{a}t Berlin} (\bibinfo{year}{2017}).

\bibitem[{\citenamefont{Langhoff and Bauschlicher}({1995})}]{Langhoff.95.calc}
\bibinfo{author}{\bibfnamefont{S.}~\bibnamefont{Langhoff}} \bibnamefont{and}
  \bibinfo{author}{\bibfnamefont{J.}~\bibnamefont{Bauschlicher}},
  \bibinfo{journal}{J. Chem. Phys.} \textbf{\bibinfo{volume}{{102}}},
  \bibinfo{pages}{5220} (\bibinfo{year}{{1995}}).

\bibitem[{\citenamefont{Schwenke and Partridge}({2000})}]{Schwenke.00.calc}
\bibinfo{author}{\bibfnamefont{D.~W.} \bibnamefont{Schwenke}} \bibnamefont{and}
  \bibinfo{author}{\bibfnamefont{H.}~\bibnamefont{Partridge}},
  \bibinfo{journal}{J. Chem. Phys.} \textbf{\bibinfo{volume}{{113}}},
  \bibinfo{pages}{6592} (\bibinfo{year}{{2000}}).

\bibitem[{\citenamefont{Huang et~al.}({2013})\citenamefont{Huang, Freedman,
  Tashkun, Schwenke, and Lee}}]{Huang.13.calc}
\bibinfo{author}{\bibfnamefont{X.}~\bibnamefont{Huang}},
  \bibinfo{author}{\bibfnamefont{R.}~\bibnamefont{Freedman}},
  \bibinfo{author}{\bibfnamefont{S.}~\bibnamefont{Tashkun}},
  \bibinfo{author}{\bibfnamefont{D.}~\bibnamefont{Schwenke}}, \bibnamefont{and}
  \bibinfo{author}{\bibfnamefont{T.}~\bibnamefont{Lee}}, \bibinfo{journal}{J.
  Quant. Spectrosc. Radiat. Transf.} \textbf{\bibinfo{volume}{{130}}},
  \bibinfo{pages}{134–46} (\bibinfo{year}{{2013}}).

\bibitem[{\citenamefont{Nikitin et~al.}({2013})\citenamefont{Nikitin, Rey, and
  Tyuterev}}]{Nikitin.13.calc}
\bibinfo{author}{\bibfnamefont{A.}~\bibnamefont{Nikitin}},
  \bibinfo{author}{\bibfnamefont{M.}~\bibnamefont{Rey}}, \bibnamefont{and}
  \bibinfo{author}{\bibfnamefont{V.}~\bibnamefont{Tyuterev}},
  \bibinfo{journal}{Chem. Phys. Lett.} \textbf{\bibinfo{volume}{{565}}},
  \bibinfo{pages}{5} (\bibinfo{year}{{2013}}).

\bibitem[{\citenamefont{Li et~al.}({2015})\citenamefont{Li, Gordon, Rothman,
  Tan, Hu, Kassi, Campargue, and Medvedev}}]{Li.15.calc}
\bibinfo{author}{\bibfnamefont{G.}~\bibnamefont{Li}},
  \bibinfo{author}{\bibfnamefont{I.}~\bibnamefont{Gordon}},
  \bibinfo{author}{\bibfnamefont{L.}~\bibnamefont{Rothman}},
  \bibinfo{author}{\bibfnamefont{Y.}~\bibnamefont{Tan}},
  \bibinfo{author}{\bibfnamefont{S.-M.} \bibnamefont{Hu}},
  \bibinfo{author}{\bibfnamefont{S.}~\bibnamefont{Kassi}},
  \bibinfo{author}{\bibfnamefont{A.}~\bibnamefont{Campargue}},
  \bibnamefont{and} \bibinfo{author}{\bibfnamefont{E.}~\bibnamefont{Medvedev}},
  \bibinfo{journal}{Astrophys J. Supplement Series}
  \textbf{\bibinfo{volume}{{216}}}, \bibinfo{pages}{15}
  (\bibinfo{year}{{2015}}).

\bibitem[{\citenamefont{Tyuterev et~al.}(2019)\citenamefont{Tyuterev, Barbe,
  Jacquemart, Janssen, Mikhailenko, and Starikova}}]{Tyut.19.calc}
\bibinfo{author}{\bibfnamefont{V.}~\bibnamefont{Tyuterev}},
  \bibinfo{author}{\bibfnamefont{A.}~\bibnamefont{Barbe}},
  \bibinfo{author}{\bibfnamefont{D.}~\bibnamefont{Jacquemart}},
  \bibinfo{author}{\bibfnamefont{C.}~\bibnamefont{Janssen}},
  \bibinfo{author}{\bibfnamefont{S.}~\bibnamefont{Mikhailenko}},
  \bibnamefont{and}
  \bibinfo{author}{\bibfnamefont{E.}~\bibnamefont{Starikova}},
  \bibinfo{journal}{J. Chem. Phys.} \textbf{\bibinfo{volume}{150}},
  \bibinfo{pages}{184303} (\bibinfo{year}{2019}).

\bibitem[{\citenamefont{Chang et~al.}({2019})\citenamefont{Chang, Guo, Wang,
  Mou, Ren, Ma, and Guo}}]{Chang.19.calc}
\bibinfo{author}{\bibfnamefont{J.}~\bibnamefont{Chang}},
  \bibinfo{author}{\bibfnamefont{L.}~\bibnamefont{Guo}},
  \bibinfo{author}{\bibfnamefont{R.}~\bibnamefont{Wang}},
  \bibinfo{author}{\bibfnamefont{J.}~\bibnamefont{Mou}},
  \bibinfo{author}{\bibfnamefont{H.}~\bibnamefont{Ren}},
  \bibinfo{author}{\bibfnamefont{J.}~\bibnamefont{Ma}}, \bibnamefont{and}
  \bibinfo{author}{\bibfnamefont{H.}~\bibnamefont{Guo}}, \bibinfo{journal}{J.
  Phys. Chem. A} \textbf{\bibinfo{volume}{{123}}}, \bibinfo{pages}{4232}
  (\bibinfo{year}{{2019}}).

\bibitem[{\citenamefont{Mizus et~al.}(2018)\citenamefont{Mizus, Kyuberis,
  Zobov, Makhnev, Polyansky, and Tennyson}}]{jt714}
\bibinfo{author}{\bibfnamefont{I.~I.} \bibnamefont{Mizus}},
  \bibinfo{author}{\bibfnamefont{A.~A.} \bibnamefont{Kyuberis}},
  \bibinfo{author}{\bibfnamefont{N.~F.} \bibnamefont{Zobov}},
  \bibinfo{author}{\bibfnamefont{V.~Y.} \bibnamefont{Makhnev}},
  \bibinfo{author}{\bibfnamefont{O.~L.} \bibnamefont{Polyansky}},
  \bibnamefont{and} \bibinfo{author}{\bibfnamefont{J.}~\bibnamefont{Tennyson}},
  \bibinfo{journal}{Phil. Trans. Royal Soc. London A}
  \textbf{\bibinfo{volume}{376}}, \bibinfo{pages}{20170149}
  (\bibinfo{year}{2018}).

\bibitem[{\citenamefont{Schwenke}(2001)}]{01Schwenke.H2O}
\bibinfo{author}{\bibfnamefont{D.~W.} \bibnamefont{Schwenke}},
  \bibinfo{journal}{J. Phys. Chem. A} \textbf{\bibinfo{volume}{105}},
  \bibinfo{pages}{2352} (\bibinfo{year}{2001}).

\bibitem[{\citenamefont{Shirin et~al.}(2003)\citenamefont{Shirin, Polyansky,
  Zobov, Barletta, and Tennyson}}]{jt308}
\bibinfo{author}{\bibfnamefont{S.~V.} \bibnamefont{Shirin}},
  \bibinfo{author}{\bibfnamefont{O.~L.} \bibnamefont{Polyansky}},
  \bibinfo{author}{\bibfnamefont{N.~F.} \bibnamefont{Zobov}},
  \bibinfo{author}{\bibfnamefont{P.}~\bibnamefont{Barletta}}, \bibnamefont{and}
  \bibinfo{author}{\bibfnamefont{J.}~\bibnamefont{Tennyson}},
  \bibinfo{journal}{J. Chem. Phys.} \textbf{\bibinfo{volume}{118}},
  \bibinfo{pages}{2124} (\bibinfo{year}{2003}).

\bibitem[{\citenamefont{Shirin et~al.}(2008)\citenamefont{Shirin, Zobov,
  Ovsyannikov, Polyansky, and Tennyson}}]{jt438}
\bibinfo{author}{\bibfnamefont{S.~V.} \bibnamefont{Shirin}},
  \bibinfo{author}{\bibfnamefont{N.~F.} \bibnamefont{Zobov}},
  \bibinfo{author}{\bibfnamefont{R.~I.} \bibnamefont{Ovsyannikov}},
  \bibinfo{author}{\bibfnamefont{O.~L.} \bibnamefont{Polyansky}},
  \bibnamefont{and} \bibinfo{author}{\bibfnamefont{J.}~\bibnamefont{Tennyson}},
  \bibinfo{journal}{J. Chem. Phys.} \textbf{\bibinfo{volume}{128}},
  \bibinfo{pages}{224306} (\bibinfo{year}{2008}).

\bibitem[{\citenamefont{Bubukina et~al.}(2011)\citenamefont{Bubukina, Zobov,
  Polyansky, Shirin, and Yurchenko}}]{Bubu.11.calc}
\bibinfo{author}{\bibfnamefont{I.}~\bibnamefont{Bubukina}},
  \bibinfo{author}{\bibfnamefont{N.}~\bibnamefont{Zobov}},
  \bibinfo{author}{\bibfnamefont{O.}~\bibnamefont{Polyansky}},
  \bibinfo{author}{\bibfnamefont{S.}~\bibnamefont{Shirin}}, \bibnamefont{and}
  \bibinfo{author}{\bibfnamefont{S.}~\bibnamefont{Yurchenko}},
  \bibinfo{journal}{Opt. \&\ Spectrosc.} \textbf{\bibinfo{volume}{110}},
  \bibinfo{pages}{160–166} (\bibinfo{year}{2011}).

\bibitem[{\citenamefont{Polyansky et~al.}(2018)\citenamefont{Polyansky,
  Kyuberis, Zobov, Tennyson, Yurchenko, and Lodi}}]{jt734}
\bibinfo{author}{\bibfnamefont{O.~L.} \bibnamefont{Polyansky}},
  \bibinfo{author}{\bibfnamefont{A.~A.} \bibnamefont{Kyuberis}},
  \bibinfo{author}{\bibfnamefont{N.~F.} \bibnamefont{Zobov}},
  \bibinfo{author}{\bibfnamefont{J.}~\bibnamefont{Tennyson}},
  \bibinfo{author}{\bibfnamefont{S.~N.} \bibnamefont{Yurchenko}},
  \bibnamefont{and} \bibinfo{author}{\bibfnamefont{L.}~\bibnamefont{Lodi}},
  \bibinfo{journal}{Mon. Not. R. Astron. Soc.} \textbf{\bibinfo{volume}{480}},
  \bibinfo{pages}{2597} (\bibinfo{year}{2018}).

\bibitem[{\citenamefont{Furtenbacher et~al.}(2020)\citenamefont{Furtenbacher,
  T\'obi\'as, Tennyson, Polyansky, and Cs\'asz\'ar}}]{jt795}
\bibinfo{author}{\bibfnamefont{T.}~\bibnamefont{Furtenbacher}},
  \bibinfo{author}{\bibfnamefont{R.}~\bibnamefont{T\'obi\'as}},
  \bibinfo{author}{\bibfnamefont{J.}~\bibnamefont{Tennyson}},
  \bibinfo{author}{\bibfnamefont{O.~L.} \bibnamefont{Polyansky}},
  \bibnamefont{and} \bibinfo{author}{\bibfnamefont{A.~G.}
  \bibnamefont{Cs\'asz\'ar}}, \bibinfo{journal}{J. Phys. Chem. Ref. Data}
  \textbf{\bibinfo{volume}{49}}, \bibinfo{pages}{033101}
  (\bibinfo{year}{2020}).

\bibitem[{\citenamefont{Lodi and Tennyson}(2010)}]{jt475}
\bibinfo{author}{\bibfnamefont{L.}~\bibnamefont{Lodi}} \bibnamefont{and}
  \bibinfo{author}{\bibfnamefont{J.}~\bibnamefont{Tennyson}},
  \bibinfo{journal}{J. Phys. B: At. Mol. Opt. Phys.}
  \textbf{\bibinfo{volume}{43}}, \bibinfo{pages}{133001}
  (\bibinfo{year}{2010}).

\bibitem[{\citenamefont{Partridge and Schwenke}(1997)}]{97PaScxx.H2O}
\bibinfo{author}{\bibfnamefont{H.}~\bibnamefont{Partridge}} \bibnamefont{and}
  \bibinfo{author}{\bibfnamefont{D.~W.} \bibnamefont{Schwenke}},
  \bibinfo{journal}{J. Chem. Phys.} \textbf{\bibinfo{volume}{106}},
  \bibinfo{pages}{4618} (\bibinfo{year}{1997}).

\bibitem[{\citenamefont{Schwenke and Partridge}(2000)}]{00ScPaxx.H2O}
\bibinfo{author}{\bibfnamefont{D.~W.} \bibnamefont{Schwenke}} \bibnamefont{and}
  \bibinfo{author}{\bibfnamefont{H.}~\bibnamefont{Partridge}},
  \bibinfo{journal}{J. Chem. Phys.} \textbf{\bibinfo{volume}{113}},
  \bibinfo{pages}{6592} (\bibinfo{year}{2000}).

\bibitem[{\citenamefont{Lodi et~al.}(2008)\citenamefont{Lodi, Tolchenov,
  Tennyson, Lynas-Gray, Shirin, Zobov, Polyansky, {Cs\'asz\'ar}, {van Stralen},
  and Visscher}}]{jt424}
\bibinfo{author}{\bibfnamefont{L.}~\bibnamefont{Lodi}},
  \bibinfo{author}{\bibfnamefont{R.~N.} \bibnamefont{Tolchenov}},
  \bibinfo{author}{\bibfnamefont{J.}~\bibnamefont{Tennyson}},
  \bibinfo{author}{\bibfnamefont{A.~E.} \bibnamefont{Lynas-Gray}},
  \bibinfo{author}{\bibfnamefont{S.~V.} \bibnamefont{Shirin}},
  \bibinfo{author}{\bibfnamefont{N.~F.} \bibnamefont{Zobov}},
  \bibinfo{author}{\bibfnamefont{O.~L.} \bibnamefont{Polyansky}},
  \bibinfo{author}{\bibfnamefont{A.~G.} \bibnamefont{{Cs\'asz\'ar}}},
  \bibinfo{author}{\bibfnamefont{J.}~\bibnamefont{{van Stralen}}},
  \bibnamefont{and} \bibinfo{author}{\bibfnamefont{L.}~\bibnamefont{Visscher}},
  \bibinfo{journal}{J. Chem. Phys.} \textbf{\bibinfo{volume}{128}},
  \bibinfo{pages}{044304} (\bibinfo{year}{2008}).

\bibitem[{\citenamefont{Gordon et~al.}(2022)\citenamefont{Gordon, Rothman,
  Hargreaves, Hashemi, Karlovets, Skinner, Conway, Hill, Kochanov, Tan
  et~al.}}]{jt836}
\bibinfo{author}{\bibfnamefont{I.~E.} \bibnamefont{Gordon}},
  \bibinfo{author}{\bibfnamefont{L.~S.} \bibnamefont{Rothman}},
  \bibinfo{author}{\bibfnamefont{R.~J.} \bibnamefont{Hargreaves}},
  \bibinfo{author}{\bibfnamefont{R.}~\bibnamefont{Hashemi}},
  \bibinfo{author}{\bibfnamefont{E.~V.} \bibnamefont{Karlovets}},
  \bibinfo{author}{\bibfnamefont{F.~M.} \bibnamefont{Skinner}},
  \bibinfo{author}{\bibfnamefont{E.~K.} \bibnamefont{Conway}},
  \bibinfo{author}{\bibfnamefont{C.}~\bibnamefont{Hill}},
  \bibinfo{author}{\bibfnamefont{R.~V.} \bibnamefont{Kochanov}},
  \bibinfo{author}{\bibfnamefont{Y.}~\bibnamefont{Tan}}, \bibnamefont{et~al.},
  \bibinfo{journal}{J. Quant. Spectrosc. Radiat. Transf.}
  \textbf{\bibinfo{volume}{277}}, \bibinfo{pages}{107949}
  (\bibinfo{year}{2022}).

\bibitem[{\citenamefont{Vasilchenko et~al.}({2021})\citenamefont{Vasilchenko,
  Mikhailenko, and Campargue}}]{21VaMiCa.H2O}
\bibinfo{author}{\bibfnamefont{S.}~\bibnamefont{Vasilchenko}},
  \bibinfo{author}{\bibfnamefont{S.~N.} \bibnamefont{Mikhailenko}},
  \bibnamefont{and}
  \bibinfo{author}{\bibfnamefont{A.}~\bibnamefont{Campargue}},
  \bibinfo{journal}{J. Quant. Spectrosc. Radiat. Transf.}
  \textbf{\bibinfo{volume}{{275}}}, \bibinfo{pages}{107847}
  (\bibinfo{year}{{2021}}).

\end{thebibliography}

\pagebreak

\end{document}